\documentclass[runningheads]{llncs}
\usepackage[T1]{fontenc}
\usepackage[]{graphicx}
\usepackage[utf8]{inputenc}
 \usepackage{setspace}
\usepackage{enumerate}
\usepackage{textcomp}

\usepackage{color,colortbl}
\definecolor{Gray}{gray}{0.9}

\usepackage{balance}
\usepackage{array}
\usepackage[hyphens]{url}
\usepackage{todonotes}
\begin{document}
\title{Nudging Using Autonomous Agents: Risks and Ethical Considerations}
%
%
\author{Vivek Nallur\inst{1}\orcidID{0000-0003-0447-4150} \and
Karen Renaud\inst{2}\orcidID{0000-0002-7187-6531} \and
Aleksei Gudkov\inst{3}\orcidID{0000-0002-3789-3813}}
\authorrunning{V. Nallur et al.}
%
\institute{School of Computer Science, University College Dublin, Ireland \\
              \email{vivek.nallur@ucd.ie} 
\and
Computer and Information Sciences, University of Strathclyde, United Kingdom\\
              \email{karen.renaud@strath.ac.uk}
\and
Independent Researcher
            \email{ag2868@gmail.com}
}
\maketitle              
\begin{abstract}
This position paper briefly discusses nudging, its use by autonomous agents, potential risks and ethical considerations while creating such systems. Instead of taking a normative approach, which guides all situations, the paper proposes a risk-driven questions-and-answer approach. The paper takes the position that this is a pragmatic method, that is transparent about beneficial intentions, foreseeable risks, and mitigations. Given the uncertainty in AI and autonomous agent capabilities, we believe that such pragmatic methods offer a plausibly safe path, without sacrificing flexibility in domain and technology.

\keywords{Nudges  \and Ethical Considerations \and Guidelines}
\end{abstract}
%
%
%

\section{Introduction}
Almost all animals (including human beings) use heuristics in their daily decision-making processes, and these have been shown to be surprisingly effective~\cite{gigerenzer_homo_2009}. Heuristics are characterized by the frugality of information and computation required to reach a satisfactory solution. While the near-universal prevalence of heuristics is testament to their evolutionary `staying power', heuristics have also been implicated in various sub-optimal decisions made by human beings~\cite{kahneman_model_2005}. These heuristics have been categorized as cognitive biases, and more than a hundred have been catalogued, with different kinds of decisions being affected by different biases~\cite{carterBehavioralSupplyManagement2007}. A nudge, first introduced by Thaler and Sunstein \cite{thaler2021nudge}, is a manipulation of the choice architecture within which a decision is made. Created as a way to help people to make wiser decisions, a nudge leverages cognitive biases present in every human being. A cognitive bias refers to the use of mental shortcuts and heuristics in decision-making~\cite{kahneman_model_2005}. These biases result in a predictable pattern of decision-making, when presented with a given context. Although \textit{some} cognitive biases exist in every human being, not every human being is affected by the \textit{same} ones. More than a hundred biases have been documented~\cite{carter_behavioral_2007} with various kinds of decision-making being affected by various categories of biases. According to Thaler and Sunstein~\cite{thaler2003libertarian}, a nudge must only be used to make a beneficial decision more likely to be chosen (e.g., choosing the default option of investing in a pension plan), or a deleterious decision more difficult to make (e.g., move sugary or fatty foods to a higher counter). This has been termed, by critics, as `Libertarian Paternalism' \cite{thaler2003libertarian}. This is so because a nudge \textit{ought} not reduce any pre-nudge options, retaining them for the nudgee to choose if they wish to. Nudging has been deployed in a wide range of contexts including: reducing sugar intake~\cite{mikkelsen2021does}, increasing number of blood donations~\cite{whyte2012nudge} and encouraging tax compliance~\cite{andersson2023nudges}.Some have expressed concerns about their unethical use as well~\cite{Kuyer,junger2024ethics}, especially when people are unaware of the presence of the nudge~\cite{alempaki2023tainted}. Zimmermann and Renaud~\cite{zimmermann2021nudge} find that making a nudge visible does not reduce its efficacy, so there does not seem to be a good reason for them to exert their influence on unwitting nudgees. Thaler and Sunstein's \emph{raison d'\^{e}tre} is that nudges must be used for the good \emph{of the nudgee}. As with any tool, some will use them for purposes that the nudgee would not agree with~\cite{lavi2018evil,sunstein2017misconceptions}. Nudging can exert influence at different stages of political or economic processes.  The deployment of negative emotions in a nudge might well run counter to the nudge's initial intention. For example, the UK's government has been criticized for their use of fear-based nudges during the COVID pandemic~\cite{UKNudge,Hume21,dodsworth2021state}. 

\section{Technology Mediated Nudging}
Since the advent of nudging, there have been many success stories in the real world using physical nudging techniques. Technology-mediated nudging has recently become prevalent by extending physical nudging to the digital context~\cite{weinmann2016digital}. Digital nudging mainly focuses on altering elements of a device's user interface to guide a user's decision. Unlike physical nudging, a significant part of digital nudging is determining relevant psychological effects (Cognitive Biases). In digital nudging literature, there is no fixed approach to determining cognitive biases in an individual~\cite{caraban2020nudge}. It can be challenging to identify cognitive biases and how they relate to a particular decision context. However, all human beings are not affected by the same subset of biases~\cite{caraban23WaysNudge2019}. 
This means that for a nudge to be effective, the first step is to detect a cognitive bias~\cite{schneiderDigitalNudgingGuiding2018} at an individual level.

Using these biases to create a nudge may seem like a frightening (almost dystopian) concept, but in effectiveness, they are very much like advertising. This is due to the fact that not all human beings are susceptible to the same biases. That is, a manipulation of the choice architecture that works on one human, may not work for another. Although all human beings suffer from some cognitive bias or the other, creating a single nudge that affects all human beings is quite difficult, if not impossible.

However, in the age of the smartphone, there is enough data available to create a personalized profile of an individual. The number of times they check social media, the amount of time between receipt of a message and response to it, the kinds of websites they visit (e.g., health, finance, sports, news), etc. can all be used to infer which kinds of bias an individual may be most likely to have. Combine these with a bag of pre-determined nudges, and a reinforcement learning algorithm, a smartphone is now able to determine with a high degree of accuracy, which nudge is mostly likely to work on a particular individual. This is called a \textit{HyperNudge}~\cite{yeungHypernudgeBigData2016}. The HyperNudge is an adaptive and personalized form of  nudging that can mine multiple sources of information to generate the highest probability of the nudge being effective. Typical sources include:
\begin{enumerate}

\item Social Media: The smartphone is able to make note of all the social channels that you use, which in turn allows it to create a profile of whether you prefer text-based communication or image/video based communication.
\item	Sentiment Analysis: All the text that is typed in via the smartphone's keyboard can be analyzed for the prevailing sentiment/mood that the user is in
\item	Behaviour Tracking: Creating a profile of the user through the day, with regard to travel, shopping, commuting, health, finances, etc.
\end{enumerate}

All of these can be combined in creative ways to achieve an automated psychographic profile of a user. If this user matches other users, then it is likely that nudges that worked on similar users are likely to work as well. A reinforcement-learning algorithm can now be employed to attempt a plethora of nudges at different times of the day. The ones that the user is most likely to respond to, can be isolated without too much effort. 
For example, a conversational agent (such as a chatbot on a website, or personalizable ones like Alexa, Siri, or Cortana to name a few) can quite easily manipulate its own speed and response to magnify the human’s cognitive bias, and nudge the human to perform an action or take a decision. Regardless of whether the nudge was intended to be for the users' benefit or not, the fact that the adaptation was performed autonomously by the agent means that the human user lacks the agency to withdraw consent. This has implications for trust and autonomy. However, since choice-architectures (whether digital or physical) exist all around us, a nudge-less environment might not exist. All user-interfaces, colour schemes, positioning of icons, ordering of messages, are geared towards optimizing some variable or the other, and therefore present some nudges, whether intentional or not. Therefore, he question of ethical considerations while creating choice architectures, must be grappled with.

There are already attempts to detect cognitive biases via conversational agents~\cite{pilli_exploring_2023} as well as attempts to use social robots to deliver  nudges~\cite{alimehenniNudgesConversationalAgents2021}. This suggests a clear need for ethical guidelines, to inform creation of autonomous agents that nudge.

Another area of focus is the integration of artificial intelligence (AI) machines as teammates in collaborative decision-making. Seeber et al. \cite{seeber2020machines} explore the risks and benefits of AI machines as teammates, including their ability to recognize and counteract human cognitive biases. Delgado et al. \cite{delgado2018opportunities} highlight the potential use of chatbots in decision support systems for reducing energy consumption in government facilities. These studies underscore the importance of considering cognitive biases in decision-making contexts and exploring the role of AI machines and chatbots.

In the domain of recommender systems, Theocharous et al.~\cite{theocharous2019personalizing} propose the incorporation of personalized cognitive bias measurement to enhance the accuracy of recommendations. They argue that by explicitly modeling users' cognitive biases, recommender systems can present options that align more closely with users' preferences. However, further research is needed to validate these claims. 

\paragraph{Autonomous and Self-Adaptive Systems:}
Adaptive Systems (also sometimes called self-adaptive systems) focus on reconfiguring themselves in response to changing environments~\cite{song2015architectural},  ensuring fair outcomes in decentralized resource allocation ~\cite{nallur_clonal_2016}, modelling emergence\cite{nallur2016algorithm}, emotion modelling~\cite{horned_models_2023}, etc. Hence, it is not far-fetched to speculate that self-adaptive choice architectures, augmented with AI-based user profiling, will be used for nudging. We refer to such autonomous systems that use AI to target and deliver nudges as AI-driven Decision Architectures (AIDA).

\section{Examples of Biases}
There are a plethora of cognitive biases that can be targetted by a nudge~\footnote{\url{https://en.wikipedia.org/wiki/List_of_cognitive_biases}}. It is essential to note that while these are called biases, there is no consensus about whether, they should be thought of as leading to errors in decision-making. Some researchers consider these to be merely patterns in decision-making~\cite{gigerenzer_homo_2009}. Although there is no definitive list, with different researchers highlighting different sets, there are some that are consistently present in most cataloging efforts. Here we list five of the most common ones, merely to illustrate the kinds of patterns that may affect users' decision-making.

\paragraph{Anchoring and Adjustment:}
The phenomenon of anchoring and adjustment refers to the tendency of individuals to be significantly influenced by an initial value or reference point (referred to as the anchor) when making subsequent judgments or decisions. Upon encountering the anchor, individuals typically make adjustments based on it, but these adjustments often prove to be insufficient, leading to biased outcomes ~\cite{furnham2011literature}. This bias arises because individuals tend to rely on the anchor as a starting point and then adjust their judgments or estimates from there. However, these adjustments frequently fail to fully account for the actual underlying value or information, resulting in a persistent bias towards the initial anchor.

Anchoring and adjustment can exert its impact across a range of decision-making contexts, including negotiations, pricing decisions, financial judgments, and even personal choices. Recognizing the influence of anchoring and adjustment allows individuals to enhance their awareness of this bias and make more reasoned and rational decisions.

\paragraph{Availability Heuristic:}
Availability Heuristic~\cite{macleod1992memory}  refers to the tendency to rely on easily accessible examples or information when making judgments or decisions. This bias influences decision-making by prioritizing readily available information over a broader range of data. Additionally, it can affect non-decision-making processes like perceptions and evaluations. 

\paragraph{Sunk Cost:} The Sunk Cost Fallacy ~\cite{haita2017sunk} describes the inclination to persist with investments in a project or decision despite its diminishing rationality or benefit. This bias specifically pertains to decision-making, as it influences ongoing investment choices.

\paragraph{Spotlight Effect:} The Spotlight Effect ~\cite{gilovich1999spotlight}  involves overestimating the extent to which others notice or remember one's behavior, primarily relating to social perception and self-consciousness rather than directly impacting decision-making processes.

\paragraph{Framing:} The framing effect is a cognitive bias wherein the presentation or framing of information or options influences decision-making processes ~\cite{tversky1981framing}. For instance, when the same information is framed as a gain or a loss, people exhibit risk-averse behavior in the case of potential losses but display more risk-seeking behavior in the case of potential gains. This bias highlights the impact of contextual cues and information framing on decision-making, illustrating that choices can be influenced not only by option content but also by the way they are presented, indicating the presence of a decision-making bias.

\subsection{Risks of Using Biases to Nudges}
The presence of these kinds of biases indicates the potential for exploiting the fast, intuitive nature of our everyday decision-making (also called System 1 thinking), that precludes reflection. While this seems like a low-cost way to influence decision-making at scale, using nudges introduces risks as well. Since the fundamental point of the nudge is that it attempts to circumvent the users rational processes, the most obvious risk is that nudging results in a violation of an individual's autonomy~\cite{zimmermann2021nudge,kuyer2023nudge}. Apart from this, we list some less-considered risks of creating a nudge:
\begin{itemize}
    \item Triggering reactance, backfiring and having the opposite effect \cite{bicchieri2022nudging}.
    \item Presence of weak punishment for nonconformity leading to negative outcomes~\cite{bicchieri2022nudging}.
    \item Fear-based nudge leading to long term adverse effects~\cite{dodsworth2021state,kuyer2023nudge}.
    \item Undermining effectiveness of a consumer protection measure~\cite{newall2022nudge}.
    \item Infantilization of nudgees~\cite{kuyer2023nudge}.
    \item Nudge lacking legitimacy because it does not align with policy~\cite{grune2016nudge}.
    \item Causing people to mistrust choice architects~\cite{kuyer2023nudge}.
    \item Difficulties identifying optimal situations for nudging~\cite{raskoff2022nudges}.
    \item Inept or corrupt nudge designs~\cite{thalerNudgeImprovingDecisions2009}.
\end{itemize}
The presence of these (and other) risks creates an ethical dilemma. While the intentions of the \textit{nudger} may be good, the consequences for the \textit{nudgee} or wider society may not be. 

\section{Ethical Considerations}
The notion that AI embodied in physical and virtual forms has significant ethical implications for individuals and society is now widely accepted. This recognition has led to a proliferation of conferences, talks, and workshops across multiple academic disciplines dedicated to analyzing how concepts of bias, fairness, and justice can be encoded into algorithmic systems and how these systems can be evaluated from an ethical standpoint. Typically, these ethical questions, atleast in the realm of computer science are broadly categorized into two domains: interpersonal ethics (ethics-in-the-small) and the ethics of sociotechnical systems (ethics-in-the-large). It remains to be seen whether there is a clear dichotomy between these two categories. Notably, implementers of ethics in machines, robots, and other AI-enabled entities tend to primarily focus on the ethics-in-the-large domain. That is, they often turn to well-specified normative ethical frameworks, such as Utilitarianism or Kantianism, and evaluate whether their systems act in a manner deemed satisfactory within these frameworks. This pragmatic orientation implicitly accepts that morally acceptable actions need not necessarily stem from phenomenological experience or inner qualitative states. Rather moral functionalism suffices (or atleast, satisfices) and the most important aspect is whether the autonomous agent is able to reason about ethically charged situations. A salient question that arises is \textit{which normative ethical framework would be most suitable for ethics-in-the-small}? That is, how can we trust self-adaptive and autonomous software that actively attempt to influence human behavior and decision-making. What are the various considerations that an \textit{AI-Assisted Decision Architecture} (AIDA) should take into account in their conception, design, implementation and deployment lifecycle? \textbf{Note:}~Here, we do not consider far-future AI, where systems autonomously set goals and then set about achieving them. While seductive in its implications, we believe that current AI-based autonomous systems receive significant input from their stakeholders. If the ethical framework used to create such nudging system is trustworthy, then the created system will also be trustworthy. Given our need for trust in these kinds of trust, the questions (considerably summarized from \cite{nallurEthicsAIBehavioralToappear}\footnote{To appear in \textit{The Cambridge Handbook on the Law, Policy, and Regulation of Human–Robot Interaction}}) that need to be answered before designing and deploying an AIDA are as follows:
\begin{enumerate}
        \item What are the relationships that the AIDA is a part of? That is, what part does the AIDA play in the various decision points, and how does the success (or failure) of the nudge affect each stakeholder?
        \item What are the stakeholders’ preferences with regard to each relationship?
        \item What actions are needed by the AIDA to preserve each relationship, with specific reference to the decision points that the AIDA is influencing? 
        \item How does the AIDA communicate with its stakeholders, about its own checks and balances?
\end{enumerate}
Note that these questions are not the definitive list of all ethical concerns that an AIDA must contend with. Rather, they form the basic framework which can be used by any stakeholder, in the conceptualization, design, development and deployment of an AIDA. From these basic questions, we derive more specific risk-driven questions that need to be answered during different phases of the nudge lifecycle.

\subsection{Fictional Usecase}
Consider an AI-assisted electronic wallet that recommends (or highlights) financial instruments to buy or sell. There are already smartphone-based applications that aim to facilitate improved financial decision-making among non-specialist individuals through the implementation of nudging techniques. Specifically, electronic wallet applications that automatically round up fractional amounts and deposit the difference into a separate savings account. This process creates a seamless and effortless mechanism for users to accumulate small savings, which could also be achieved through self-discipline, but the combination of default rounding and segregated account management provides a frictionless approach to the savings process. The integration of AI capabilities within electronic wallet applications could further enhance the potential for improved financial decision-making among lay individuals. The AI component would be able to objectively calculate the probabilities of potential gains and losses, and subsequently provide personalized recommendations regarding the most suitable financial products. This AI-driven nudging approach could enable a greater number of individuals to achieve their modest financial objectives. The underlying premise is that the goal of this nudging mechanism is to promote the well-being and benefit of the individual being influenced, rather than any ulterior motives. On the face of it, this seems quite benign and only provides benefits to the user. However, it is not entirely clear whether high-frequency messaging and persistent prompting could cross the line into application-induced harassment. Additionally, there is also the issue of who determines the acceptable risk profiles for individual users, particularly for those who may lack sufficient financial literacy to fully comprehend the associated risks. Can the user, having given consent for such nudging once (perhaps as a default option), remove consent? How would they know if this is possible?

In the next section, we outline the specific ethical values that need to be considered, while developing an AIDA. We use the fictional use-case (just outlined) as our running exemplar.

\section{Principles for the Nudge Lifecycle}
Although there seems to be a consensus (at least in Europe) on the need for regulating A/IS to make them trustworthy~\cite{NiFhaolain2020AssessingAppetiteTrustworthiness}, the proposed new AI legislation issued by the European Commission\footnote{https://data.consilium.europa.eu/doc/document/ST-5662-2024-INIT/en/pdf} categorizes the vast majority of current AI systems as “low-risk algorithms”.  However there is  recognition that manipulation of decisions can be done using AI-enabled systems. As per Recital 16 of the EU AI Act:\begin{quote}
    The placing on the market, putting into service or use of certain AI systems with the objective to or the effect of materially distorting human behaviour, whereby significant harms, in particular having sufficiently important adverse impacts on physical, psychological health or financial interests are likely to occur, are particularly dangerous and should therefore be forbidden. 
\end{quote}
This view of behaviour-change being a risky one indicates that any system, even if autonomously adapting, should ensure that any nudge delivered has been evaluated for beneficence. More specifically, without being prescriptive about which ethical framework to follow, we propose that the system stakeholders (designers, testers, deployers, to name a few) take into account, the following principles:
\begin{enumerate}
    \item \textbf{Beneficence}: This goes beyond the \textit{do-no-harm} concept. The stakeholders must actively consider and demonstrate that the nudge conveys a distinct and measurable benefit to the nudgee
    \item \textbf{Autonomy}: The nudgee must always be able to avoid or refuse the nudge. This should not result in any express or implied artificial penalties
    \item \textbf{Justice or Fairness}: Using a bottom-up learning technique such as machine-learning may result in a system that is fundamentally, statistically biased. Regardless of the technique used, it is incumbent on the stakeholders to ensure that the nudge does not result in outcomes that are unjust or unfair.
    \item \textbf{Transparent}: There is research that shows that visibility of a nudge does not reduce its effectiveness~\cite{zimmermann2021nudge}. Given this, the stakeholders of autonomous/adaptive nudging system should ensure that the nudging behaviour is transparent to the user, and any third-party that is in guardianship role to the user.
\end{enumerate}
Given these principles, we set out a series of questions that should be asked during different stages of an autonomous nudging system. 

\subsection{Requirements Gathering}
Considering the AI-augmented electronic wallet app, the relevant questions for requirements gathering are:
\begin{enumerate}
    \item What are the social relationships that the wallet is a part of?\\
    Possible answers here include the user of the app, the financial dependents of the user, the designers and deployers of the app. Each candidate relationship a higher preference on a particular value. For example, the user of the app would be concerned about their autonomy and financial safety. The financial dependents would place a higher premium on the financial safety (a consequential view). The designers and deployers would be concerned about their reputation, as a knock-on effect, if the wallet app malfunctions. 

    \item What are the decision points at which the wallet nudges? Which particular bias is targetted at which particular decision point? How do these impact the relationships identified previously? 

    \item How does the wallet app commuicate its commitment to the four principles outlined above?

\end{enumerate}

\subsection{Design Phase}
Renaud and Zimmermann~\cite{renaud2018ethical} argue that ethical considerations ought to be front and centre in nudge design. Schmidt and Engelen~\cite{schmidt2020ethics} consider arguments both for and against the use of nudges, and conclude with an admonition for `\emph{ethical assessment of nudges on a case-by-case basis}'. Given this, the design of the nudge should have the following properties: \cite{thaler2021nudge,sunstein2023eight}:
\begin{enumerate}
    \item \textbf{Always facilitate choice}:
In any given situation, the nudge must allow for the possibility of choice, \textit{i.e.}, the nudgee must be able to side-step the nudge without any effort~\cite{sunstein2014nudging}
    \item \textbf{Informed consent}:
The nudgee must be informed about the nudge and be able to give, deny or withdraw consent to being nudged
\item \textbf{No economic incentives}:
The nudgee should neither be presented with economic incentives to accept the nudge, nor any disincentives on side-stepping it. This includes extra effort being required to choose any particular option that the nudger considers undesirable. 
    \item  \textbf{Pre-Nudge Conditions Retained}:
The act of presenting a nudge should not remove any options that existed before the nudge 
    \item \textbf{Nudge for good}:
Intentional harm is prohibited. The nudge should be avoidable without affecting the system's functioning, if harm is detected at any juncture.
    \item  \textbf{Monitoring}
It is the responsibility of the nudge architect to ensure that the nudge system is continuously monitored to detect unanticipated outcomes.
    \item  \textbf{Feedback}
It falls upon the architect to provide the user with a means of obtaining feedback on the choices they have been making while using the system, so that they can be aware of the power of the nudge.
\end{enumerate}

Considering the properties that we would like a nudge to have, the following questions ought to be asked during the design phase of the AIDA wallet:
\begin{enumerate}
    \item What are the social relationships the AIDA is a part of? And how does it help to maintain them?\\
    Possible answers here may be similar to the ones derived during the requirements phase. However, ensuring that the properties outlined above (facilitating choice, informed consent, providing feedback) are maintained for all the relationship participants will affect the design choices made, in the biases and design elements chosen for the nudge.

    \item How does the AIDA communicate its intentions, checks and balances with regard to the properties of \textit{informed consent}, \textit{pre-nudge conditions retained}, \textit{nudge for good}? How does each stakeholder communicate \emph{back} to the AIDA, their own preferences with regards to these properties? How is the monitoring implemented?\\
    Again, ensuring the properties of clear and transparent communication, both from the AIDA to the nudgee and the nudgee back to the AIDA, will greatly affect the design.
    
\end{enumerate}    

\subsection{Nudge Deployment and Evaluation}
It is understood to be a basic principle of pharmacotherapy that all drugs have harmful and beneficial effects and that it is the job of the doctor to balance these, while treating a patient. While the ethics regarding  treatment of patients are well-understood, it is less clear what ethics should apply when creating an intervention that scales from a one-on-one doctor-patient relationship, to a choiceArchitect-user relationship. Not only is there a lack of direct feedback, but the temporal gap between intervention and feedback could also be too large to ensure effective remedies in case of harm. In this case, should the balance of benefit-to-harm be so hugely tilted in favour of the benefit, that to not make the intervention be viewed as a harm itself? Or would a `relatively benign' harm not require a `strong benefit'? A designer of an autonomous system might not be in a position to actively explicate the possible harm, because of the adaptive nature of the AIDA precludes systematic analysis (or makes it infeasible).

Given this context, the following questions need to be asked during the deployment phase of the AIDA wallet:
\begin{enumerate}
    \item What relationships does the large-scale deployment of the AIDA create, and participate in?\\
    Again, the stakeholders may largely be the same during this phase, but the activities needed to achieve the strong benefit, in the presence of a delayed feedback (or absence of direct feedback) are quite different. The deployment process creates a distance between the designers/implementers of the AIDA, and creates larger chains of deployment relationships. Each of these must be empowered to convey feedback promptly and completely back to the design-and-implementation team.

    \item Since many decisions involve making a tradeoff between priorities, which values will be preferred and which ones de-prioritized? Which relationships do they affect?
    \item Which values are most immediately affected by the lack of a direct feedback loop, as well as the temporal distance between the location of the nudge, and the deployment team?
\end{enumerate}
Answers to all these questions will impact the deployment strategy, in terms of how long the deployment chain is, as well as how quickly feedback must propagate along this chain.

\paragraph{Evaluating Nudges:}
We do not claim that using these questions will result in an AIDA-driven nudge, that is completely beneficial in all cases, to all participants. Mills and Whittle~\cite{mills2023seeing} propose a `4S' framework highlighting the importance of nudges being: \textbf{s}ufficient, \textbf{s}calable,  \textbf{s}ubjective, and statistically \textbf{s}ignificant. While this framework serves as a guide to creating the \textit{minimum} conditions for good AIDA-driven nudge, but they do not form a complete guide to being ethical. The process of \textit{being ethical} is a negotiated equilibrium that requires a reflexive consideration of the `Other', especially weaker participants in the relationship. This might mean being less efficient than strictly possible in nudge design or delivery, if it leads to greater availability of choice (autonomy), or decreasing the frequency of the nudge, if it leads to better and more effective monitoring. 
The evaluation of whether an AIDA-driven nudge is successful is therefore a socio-technical evaluation rather than a purely technical, objective one.


\section{Conclusion}
In this paper, we introduced some of the biases that are targetted in nudge-design, some of the risks associated with exploiting these biases. We also introduce a risk-inspired question-framework that could aid AIDA designers to proceed ethically in building nudges into their system. These questions must be asked over-and-above any values that the AIDA purports to uphold. 

\section*{Acknowledgements}
The authors wish to acknowledge the long and continuing conversations they have had with members of the IEEE P7008 group, involved with the creating a standard for ethical nudging. This authors are members of this group. However, none of the views and positions in this paper should be construed as the position of the group, or the IEEE. All views are the authors' own.

%
%
\bibliographystyle{splncs04}
\bibliography{refs}
\end{document}